\documentclass[12pt]{iopart}
\newcommand{\gguide}
\usepackage[dvips]{graphicx}
\usepackage{rotating} 

\begin{document}

\title[]{Which cosmological model --- with dark energy or modified FRW dynamics?}

\author{Marek Szyd{\l}owski}
\address{Astronomical Observatory, Jagiellonian University,
Orla 171, 30-244 Krak{\'o}w, Poland}
\address{Complex Systems Research Centre, Jagiellonian University,
Reymonta 4, 30-059 Krak{\'o}w, Poland}
\ead{szydlo@oa.uj.edu.pl}

\author{W{\l}odzimierz God{\l}owski}
\address{Astronomical Observatory, Jagiellonian University,
Orla 171, 30-244 Krak{\'o}w, Poland}


\begin{abstract}
Recent measurements of distant type Ia supernovae (SNIa) as well as other
observations indicate that our universe is in accelerating phase of expansion.
In principle there are two alternative explanation for such an acceleration.
While in the first approach an unknown form of energy violating the strong
energy condition is postulated, in second one some modification of FRW
dynamics is postulated. The both approaches are in well agreement with present
day observations which is the manifestation of the degeneracy problem
appearing in observational cosmology. We use the Akaike (AIC) and Bayesian
(BIC) information criteria of model selection to overcome this degeneracy and
to determine a model with such a set of parameters which gives the most
preferred fit to the SNIa data. We consider five representative evolutional
scenarios in each of groups. Among dark energy proposal the $\Lambda$CDM model,
CDM model with phantom field, CDM model with topological defect, model with
Chaplygin gas, and the model with the linear dynamical equation of  state
parameter. As an alternative prototype scenarios we consider: brane world Dvali 
Gabadadze Porrati scenario, brane models in Randall-Sundrum scenario, Cardassian 
models with dust matter and radiation, bouncing model with the cosmological 
constant and metric-affine gravity (MAG) inspired cosmological models. Applying
the model selection criteria we show that both AIC and BIC indicates that
additional contribution arises from nonstandard FRW dynamics are not necessary
to explain SNIa. Adopting the model selection information criteria we show
that the AIC indicates the flat phantom model while BIC indicates both flat
phantom and flat $\Lambda$CDM models.
\end{abstract}
\maketitle

\section{Introduction}
 
If we assume that the FRW model with a source in the form of perfect fluid
well describes the present evolution of our Universe, then there is only one
way of explain the observational fact that the Universe is accelerating
\cite{Riess:1998cb,Perlmutter:1998np,Riess:2004nr} --- to postulate some
gravity source of unknown nature which violates the strong energy condition
$\rho_{X}+3p_{X}>0$ where $\rho_{X}$ and $p_{X}$ are energy density and
pressure of dark energy, respectively \cite{Sahni:2002kh,Lahav:2004iy}. These
different candidates for dark energy were confronted with observations
\cite{Kamenshchik:2001cp,Zhu:2003sq,Sen:2002ss,Godlowski:2003pd,%
Godlowski:2004pt,Puetzfeld:2004df,Biesiada:2004td,Makler:2002jv,Avelino:2002fj,%
Bertolami:2004ic,Zhu:2004aq,Padmanabhan:2002vv,Choudhury:2003tj}.
While the most natural candidate for such a type of dark energy is vacuum
energy (the cosmological constant $\Lambda$), we are still looking for other
alternatives because the fine tuning problem with present value of the
cosmological constant. In this context it is considered an idea of dynamical
form of the equation of state (EOS) or decaying vacuum. As a first
approximation one can consider the coefficient in the EOS
$w_X \equiv \frac{p_X}{\rho_X}$ in the form linearized around the present
epoch with respect to redshift $z$ or the scale factor $a$: $w(z)=w_0+w_1z$
(or $w(z)=w_0+w_1(1-a)$, where $a=1$ corresponds to the present epoch).
The models with dust matter and such a form of the EOS for dark energy we
called the dynamical EOS (DEOS) model. In the special case when $w_1=0$ and
$w_0<-1$ we obtain the CDM quintessence model or phantom CDM (PhCDM) model.
Another interesting possibility of description dark energy offers a conception
of description of dark energy in terms of Chaplygin gas
\cite{Kamenshchik:2001cp}. In this model the equation of state has a form
$p_{X}=-\frac{A}{\rho_{X}^{\alpha}}$.
 
If we consider the FRW dynamics in which dark energy is present, the basic
equation determining the evolution has the following form
\begin{equation}
\label{eq:1}
H^{2} = \frac{\rho_{eff}}{3}-\frac{k}{a^2}.
\end{equation}
where $\rho_{eff}(a)$ is effective energy density of noninteracting
``fluids'', $k = \pm 1,0$ is the curvature index. Equation~(\ref{eq:1}) can
be presented in terms of density parameters
\begin{equation}
\label{eq:2}
\frac{H^{2}}{H_{0}^{2}} = \Omega_{eff} (z) + \Omega_{k,0}(1+z)^2
\end{equation}
where $\frac{a}{a_0}=\frac{1}{1+z}$,
$\Omega_{eff}(z)=\Omega_{m,0}(1+z)^3+\Omega_{X,0}f(z)$ and
$\Omega_{m,0}$ is the density parameter for the (baryonic and dark)
matter scaling like $a^{-3}$. For $a=a_{0}$ (the present value of the scale
factor) we obtain the constraint $\Omega_{eff,0} + \Omega_{k,0}=1$.
 
We assumed that energy density satisfies the conservation condition
\begin{equation}
\label{eq:3}
\dot{\rho}_{i} = -3H(\rho_{i} + p_{i}),
\end{equation}
for each component of the fluid $\rho_{eff}=\Sigma\rho_i$. Then from
eq.~(\ref{eq:2}) we obtain the constraint relation $\Sigma_i\Omega_{i,0}
+\Omega_{k,0}=1$.
 
All mentioned before directions which are coming toward to the description
of dark energy in the framework of standard FRW cosmology can be treated as
a representative approaches of explanation of the current Universe in terms
of dark energy. In Table~\ref{tab:1} we complete all these models together
with the dependence of Hubble's function $H=\frac{\dot{a}}{a}$ with respect
to the redshift. We also denote the number of a model's free parameters by
$d$. Note that for the flat model, $\Omega_{k,0}=0$, the number of the model
parameters is equal $d-1$. As the reference model we consider the flat FRW
model with $\Lambda=0$ (the Einstein-de Sitter model with
$\Omega_{m,0}=1$).

\begin{sidewaystable}
\begin{tabular}{c|p{3.7cm}|llc}
\hline
case & name of model    & $\qquad H(z)$& free parameters & $d$ \\ \hline
0  & Einstein-de Sitter & $H=H_0 \sqrt{\Omega_{m,0}(1+z)^3+\Omega_{k,0}(1+z)^2}$ & $H_0,\Omega_{m,0}$ & 2 \\
1  & $\Lambda$CDM       & $H=H_0 \sqrt{\Omega_{m,0}(1+z)^3+\Omega_{k,0}(1+z)^2+\Omega_{\Lambda}}$ & $H_0,\Omega_{m,0},\Omega_{\Lambda}$ & 3 \\
2  & TDCDM              & $H=H_0 \sqrt{\Omega_{m,0}(1+z)^3+\Omega_{k,0}(1+z)^2+\Omega_{T,0}(1+z)}$ & $H_0,\Omega_{m,0},\Omega_{T,0}$ & 3 \\
3a & PhCDM, $w=-\frac{4}{3}$ & $H=H_0 \sqrt{\Omega_{m,0}(1+z)^3+\Omega_{k,0}(1+z)^2+\Omega_{Ph,0}(1+z)^{3\left(1+w\right)}}$ & $H_0,\Omega_{m,0},\Omega_{Ph,0}$ & 3 \\
3b & PhCDM, $w$ -- fitted  &                                                                                               & $H_0,\Omega_{m,0},\Omega_{Ph,0},w$ & 4 \\
4a & ChGM,  $\alpha=1$           & $H=H_0 \sqrt{\Omega_{m,0}(1+z)^3+\Omega_{k,0}(1+z)^2+\Omega_{Ch,0}\left(A_s+\left(1-A_s \right)(1+z)^{3\left(1+\alpha\right)}\right)^{\frac{1}{1+\alpha}}}$ & $H_0,\Omega_{m,0},\Omega_{Ch,0},A_s$ & 4 \\
4b & GChGM, $\alpha$ -- fitted   &                                                                                        & $H_0,\Omega_{m,0},\Omega_{Ch,0},A_s,\alpha$ & 5 \\
5a & $DEQS$, $p_X=(w_0+w_1z)\rho_X$ &
     $H=H_0 \sqrt{\Omega_{m,0}(1+z)^3+\Omega_{k,0}(1+z)^2+\Omega_{X,0}(1+z)^{3\left(w_0-w_1+1\right)}e^{3w_1z}}$ & $H_0,\Omega_{m,0},\Omega_{X,0},w_0,w_1$ & 5 \\
5b & $DEQS$, $p_X=(w_0+(1-a)w_1)\rho_X$ &
     $H=H_0 \sqrt{\Omega_{m,0}(1+z)^3+\Omega_{k,0}(1+z)^2+\Omega_{X,0}(1+z)^{3\left(w_0+w_1+1\right)}e^{\frac{-3w_1z}{1+z}}}$ & $H_0,\Omega_{m,0},\Omega_{X,0},w0,w_1$ & 5 \\
\hline
\end{tabular}
\caption{The five prototypes models explaining acceleration in terms of dark
energy conceptions.}
\label{tab:1}
\end{sidewaystable}
 
Since the discovery of acceleration of the Universe the theoretical and
observational cosmology becomes in the state of permanent tension because
of opposite aims of investigations. While the theoretical investigations
go towards to generalization degree of consideration by adding some new model
parameters, the observational cosmology tries to constraint these parameters.
The main goal of observational cosmology is to find a model with a minimal
number of essential model parameters. Then this, (of course it may be a
na{\"i}ve model) is a starting point for the further analysis of constraints
from the observational data. In the present observational cosmology such
a role plays the concordance $\Lambda$CDM model and the PhCDM model
\cite{Caldwell:1999ew,Carroll:2003st,Hsu:2004vr,Godlowski:2005tw}.
 
Because nature of dark energy is unknown, it is considered another theoretical
possibility that the phenomenon of accelerated expansion is actually a sign of
a breakdown of the classical Friedmann equation which governs the expansion
rate. In this context dark energy effects can be manifestation of a
modification to the FRW equation arising from the new exotic physics.

\begin{sidewaystable}
\begin{tabular}{c|p{3.7cm}|llc}
\hline
case & name of model    & $\qquad H(z)$& free parameters & $d$ \\ \hline
1  & DGP  & $H=H_0 \sqrt{\left(\sqrt{\Omega_{m,0}(1+z)^3+\Omega_{rc,0}}+\sqrt{\Omega_{rc,0}} \right)^2+\Omega_{k,0}(1+z)^2}$ & $H_0,\Omega_{m,0},\Omega_{rc,0}$ & 3 \\
2a & RSB, $\Lambda=0$      & $H=H_0 \sqrt{\Omega_{m,0}(1+z)^3+\Omega_{k,0}(1+z)^2+\Omega_{dr,0}(1+z)^4+\Omega_{\lambda,0}(1+z)^6}$ & $H_0,\Omega_{m,0},\Omega_{\lambda,0},\Omega_{dr,0}$ & 4 \\
2b & RSB, $\Lambda \ne 0$  & $H=H_0 \sqrt{\Omega_{m,0}(1+z)^3+\Omega_{k,0}(1+z)^2+\Omega_{dr,0}(1+z)^4+\Omega_{\lambda,0}(1+z)^6+\Omega_{\Lambda}}$ & $H_0,\Omega_{m,0},\Omega_{\lambda,0},\Omega_{dr,0},\Omega_{\Lambda}$ & 5 \\
3  & Cardassian            & $H=H_0 \sqrt{\Omega_{m,0}(1+z)^3+\Omega_{r,0}(1+z)^4+\Omega_{k,0}(1+z)^2+\Omega_{CC,0}}$ & $H_0,\Omega_{m,0},\Omega_{Car,0},n$ & 4 \\
   & $(\Omega_{r,0}=0.0001)$ & $where \qquad \Omega_{CC,0} \equiv  \Omega_{Car,0}\left(\Omega_{m,0}(1+z)^3+\Omega_{r,0}(1+z)^4 \right)^n$ & &  \\
4  & B$\Lambda$CDM  & $H=H_0 \sqrt{\Omega_{m,0}(1+z)^3+\Omega_{k,0}(1+z)^2-\Omega_{n,0}(1+z)^n+\Omega_{\Lambda}}$ & $H_0,\Omega_{m,0},\Omega_{n,0},\Omega_{\Lambda},n$& 5 \\
5a & MAG $\Lambda=0$       & $H=H_0 \sqrt{\Omega_{m,0}(1+z)^3+\Omega_{k,0}(1+z)^2+\Omega_{\psi,0}(1+z)^6}$ & $H_0,\Omega_{m,0},\Omega_{\psi,0}$ & 3 \\
5b & MAG $\Lambda \ne 0$   & $H=H_0 \sqrt{\Omega_{m,0}(1+z)^3+\Omega_{k,0}(1+z)^2+\Omega_{\psi,0}(1+z)^6+\Omega_{\Lambda}}$ & $H_0,\Omega_{m,0},\Omega_{\psi,0},\Omega_{\Lambda}$ & 4 \\
\hline
\end{tabular}
\caption{The five prototypes models explaining acceleration in terms of the
modification of the FRW equation.}
\label{tab:2}
\end{sidewaystable}

According to the brane world idea, the standard particles are confined on a
hyper-surface which is called a brane, which is embedded in a higher dimensional
bulk space time in which gravity could spread \cite{Randall:2002ie}. Then some
additional contribution to the standard FRW equation may arise as a
consequence of embedding our Universe in the higher dimensional bulk space.
It is interesting that the cosmological models formulated in the framework of
brane induced gravity can explain the acceleration of the Universe without
the conception of dark energy. The additional term in the FRW equation drives
the acceleration of the Universe at a late epoch when it is dominant
\cite{Randall:1999ee,Randall:1999vf,Dvali:2000hr,Dvali:2000xg,Deffayet:2000uy}.
Therefore the effect arising from existence of these additional dimensions
can mimic dark energy through a modified FRW dynamics. As a prototype of
evolutional scenarios arising from possible extra dimensions we consider two
models: the Dvali Gabadadze Porrati (DGP) model and the Randall-Sundrum
scenario of the brane-world (RSB) model. In the DGP model there
is present a certain crossover scale $r_c$ that defines what kind of gravity
an observer located on the brane observes. While for shorter then $r_c$
distances observer measures the standard gravitational force, for larger
then $r_c$ distances gravitational force behaves like $r^{-3}$. In the RSB
models there are present some additional parameters which are absent in the
standard cosmology, namely brane tension $\lambda$ and dark radiation $U$
(see Table \ref{tab:2}). If we assume the quantum nature of the Universe then
quantum gravity corrections are important at both big bang and big rip
singularities. Note that a big rip singularities can not be generic future
state in phantom cosmology. Only if energy density is unbounded function of
time one obtains a big rip/smash \cite{McInnes:2002}.
To account of quantum effects allows to avoid the initial
singularity which phenomenologically can be modelled by a bounce
\cite{Bojowald:2001xe,Bojowald:2002nz,Nojiri:2003jn,Elizalde:2004mq,Nojiri:2004pf}.
We consider bouncing models as a prototype of evolutional scenario in which
quantum effect was important in its very early evolution.
 
Freese and Lewis \cite{Freese:2002sq} have recently proposed to modify the
FRW equation by adding a priori an additional term proportional to
$\rho_{eff}^n$ (dubbed the Cardassian term by the authors).
If we consider a single fluid with energy density $\rho$ then the Cardassian
model stays equivalent to a two component noninteracting model. In the special
case of dust matter there is a simple interpretation for the origin of this
new ``Cardassian term''---a perfect fluid satisfying the equation of state
$p=(n-1)\rho$. Therefore, if $n<0$ the phantom cosmological models can be
recovered. In our analysis we consider two fluids, matter and radiation.
However, note that the radiation term is presently small with comparison to
the matter term. We can assume $\Omega_{r,0}=0.0001$ for radiation matter
\cite{Vishwakarma:2002ek}.
 
The next possibility is the MAG cosmological model based on a non-Riemannian
gravity theory \cite{Puetzfeld:2004sw,Krawiec:2005jj}. Because this model
unifies both the RSB model with vanishing dark radiation and the model with
spinning fluid \cite{Szydlowski:2003nv} we include this model to the class of
models which are based on the modified FRW equation. Because the RSB and MAG
models without the cosmological constant do not fit well the SNIa data
\cite{Godlowski:2004pt,Krawiec:2005jj,Dabrowski:2002tp} we also analyze the
MAG and RSB models with the non-vanishing $\Lambda$ term. The five models
based on the modified FRW equation are presented in Table \ref{tab:2}.
 
Main goal of this paper is comparison of two conceptually different classes
of models using the information criteria of models selection
\cite{Liddle:2004nh}. The Bayesian information criterion gives important
information whether additional parameters introduced by ``new physics'' are
actually relevant and to have impact on the current Universe.

\section{Distant supernovae as cosmological probes dark energy or the modified FRW equation}
 
Distant type Ia supernovae surveys allowed us to find that the present Universe
is accelerating \cite{Riess:1998cb,Perlmutter:1998np}. Every year new SNIa
enlarge available data sets with more distant objects and lower systematics
errors. Riess et al. \cite{Riess:2004nr} compiled the latest samples which
become the standard data sets of SNIa. One of them, the restricted ``Gold''
sample of 157 SNIa, is used in our analysis.
 
For the distant SNIa one can directly observe their the apparent magnitude $m$
and redshift $z$. Because the absolute magnitude $\mathcal{M}$ of the
supernovae is related to its absolute luminosity $L$, then the relation
between the luminosity distance $d_L$ and the observed magnitude $m$ and
the absolute magnitude $M$ has the following form
\begin{equation}
\label{eq:4}
m - M = 5\log_{10}d_{L} + 25.
\end{equation}
Instead of $d_L$, the dimensionless parameter $D_L$
\begin{equation}
\label{eq:5}
D_{L}=H_{0}d_{L}
\end{equation}
is usually used and then eq.~(\ref{eq:4}) changes to
\begin{equation}
\label{eq:6}
\mu \equiv m - M = 5\log_{10}D_{L} + \mathcal{M}
\end{equation}
where
\begin{equation}
\label{eq:7}
\mathcal{M} = - 5\log_{10}H_{0} + 25.
\end{equation}
 
We know the absolute magnitude of SNIa from the light curve. The luminosity
distance of supernova can be obtain as the function of redshift
\begin{equation}
\label{eq:8}
d_L(z) =  (1+z) \frac{c}{H_0} \frac{1}{\sqrt{|\Omega_{k,0}|}}
\mathcal{F} \left( H_0 \sqrt{|\Omega_{k,0}|} \int_0^z \frac{d z'}{H(z')} \right)
\end{equation}
where $\Omega_{k,0} = - \frac{k}{H_0^2}$ and
\begin{equation}
\label{eq:9}
\mathcal{F} (x) = \cases { sinh(x) &for  $k<0$\\
x&for  $k=0$\\
sin(x)  &for  $k>0$\\}
\end{equation}
 
Finally it is possible to probe dark energy which constitutes the main
contribution to the matter content. It is assumed that supernovae
measurements come with the uncorrelated Gaussian errors and in this case the
likelihood function $\mathcal{L}$ can be determined from the chi-square
statistic $\mathcal{L}\propto \exp(-\chi^{2}/2)$ where
\begin{equation}
\label{eq:10}
\chi^{2}=\sum_{i}\frac{(\mu_{i}^{theor}-\mu_{i}^{obs})^{2}}
{\sigma_{i}^{2}},
\end{equation}
while the probability density function of cosmological parameters is
derived from Bayes' theorem \cite{Riess:1998cb}. Therefore, we can perform the
estimation of model parameters using the minimization procedure, based on
the likelihood function.
 
In the modern observational cosmology there is present so called the degeneracy
problem: many models with dramatically different scenarios are in good
agreement with the present data observations. Information criteria of the
model selection \cite{Liddle:2004nh} can be used to solve this degeneracy
among some subclass of dark energy models. Among these criteria the Akaike
information (AIC) \cite{Akaike:1974} and the Bayesian information criteria
(BIC) \cite{Schwarz:1978} are most popular. From these criteria we can
determine the number of the essential model parameters providing the
preferred fit to the data.
 
The AIC is defined in the following way
\cite{Akaike:1974}
\begin{equation}
\label{eq:11}
AIC = - 2\ln{\mathcal{L}} + 2d
\end{equation}
where $\mathcal{L}$ is the maximum likelihood and $d$ is a number of the
model parameters. The best model with a parameter set providing the preferred
fit to the data is that minimizes the AIC.
 
The BIC introduced by Schwarz \cite{Schwarz:1978} is defined as
\begin{equation}
\label{eq:12}
BIC = - 2\ln{\mathcal{L}} + d\ln{N}
\end{equation}
where $N$ is the number of data points used in the fit. While AIC tends to 
favour models with large number of parameters, the BIC the more strongly
penalizes them, so BIC provides the useful approximation to full evidence 
in the case of no prior on the set of model parameters \cite{Parkinson:2005}.

The effectiveness of using these criteria in the current cosmological
applications has been recently demonstrated by Liddle \cite{Liddle:2004nh}
who, taking CMB WMAP data \cite{Bennett:2003bz}, found the number of essential
cosmological parameters to be five. Moreover he obtained the important
conclusion that spatially-flat models are statistically preferred to close
models as it was indicated by the CMB WMAP analysis (their best-fit value is
$\Omega_{tot,0} \equiv \Sigma_i \Omega_{i,0} = 1.02 \pm 0.02$ at
$1\sigma$ level).

In the paper of Parkinson et~al. \cite{Parkinson:2005} the usefulness of 
Bayesian model selection criteria in the context of testing for double 
inflation with WMAP was demonstrated. These criteria was also used recently 
by us to show that models with the big-bang scenario are rather prefers over 
the bouncing scenario \cite{Szydlowski:2005qb}.

Please note that both information criteria values have no absolute sense and
only the relative values between different models are physically interesting.
For the BIC a difference of $2$ is treated as a positive evidence
($6$ as a strong evidence) against the model with larger value of BIC
\cite{Jeffreys:1961,Mukherjee:1998wp}.
Therefore one can order all models which belong to the ensemble of dark
energy models following the AIC and BIC values. If we do not find any
positive evidence from information criteria the models are treated as a
identical and eventually additional parameters are treated as not significant.
Therefore the information criteria offer the possibility of introducing
relation of weak order in the considered class of analyzed models.
 
The results of calculation of the AIC and BIC in the context of dark energy
models are presented in Tables~\ref{tab:3}-\ref{tab:6}. In Table~\ref{tab:3}
we show results for dark energy models considered for both flat and non flat
cases without any assumed extra priors, while in Table~\ref{tab:4} we
presented results for models with the modified FRW equation. In general case
the number of essential parameters in the cosmological models with dark
energy is in principal two, i.e., $H_0$, $\Omega_{m,0}$. It means that
the flat model is favored in the light of the information criteria. We can
observe two rival models which minimize the AIC and BIC. They are the
$\Lambda$CDM model and the phantom CDM (PhCDM) model. One can observe that
both BIC and AIC values assume lower values for phantom models. It can be
regarded as a positive evidence in favor of the PhCDM model. Basing on these
simple and objective information criteria we obtain that SNIa data favored
the models with the initial (big bang) and final (big rip) singularities.
 
\begin{figure}
\includegraphics[width=0.9\textwidth]{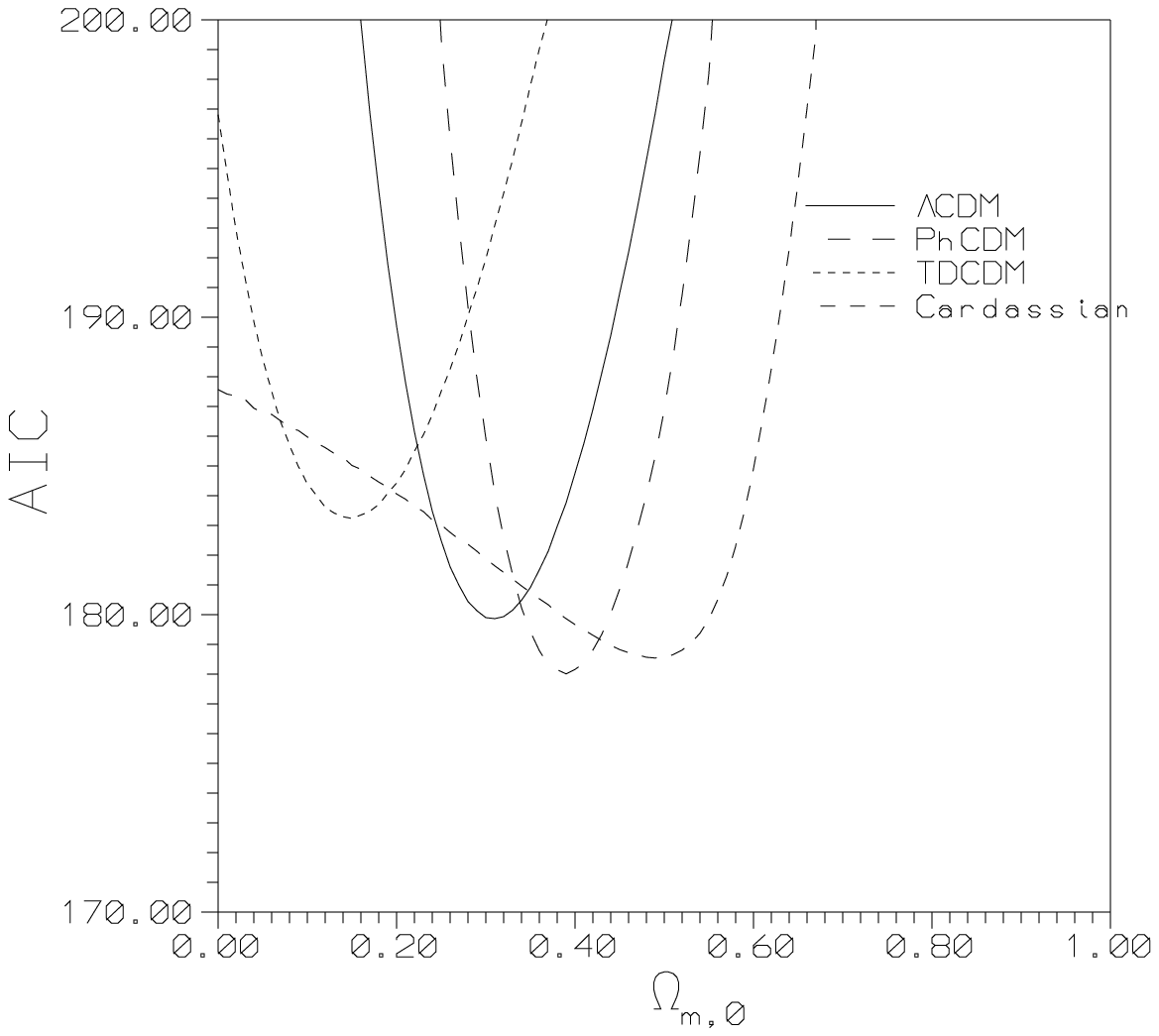}
\caption{The values of the AIC in respect to fixed value
$\Omega_{m,0}$ for four flat models (with topological defect, with
the cosmological constant, with phantom and for the Cardassian model).}
\label{fig:1}
\end{figure}

\begin{figure}
\includegraphics[width=0.9\textwidth]{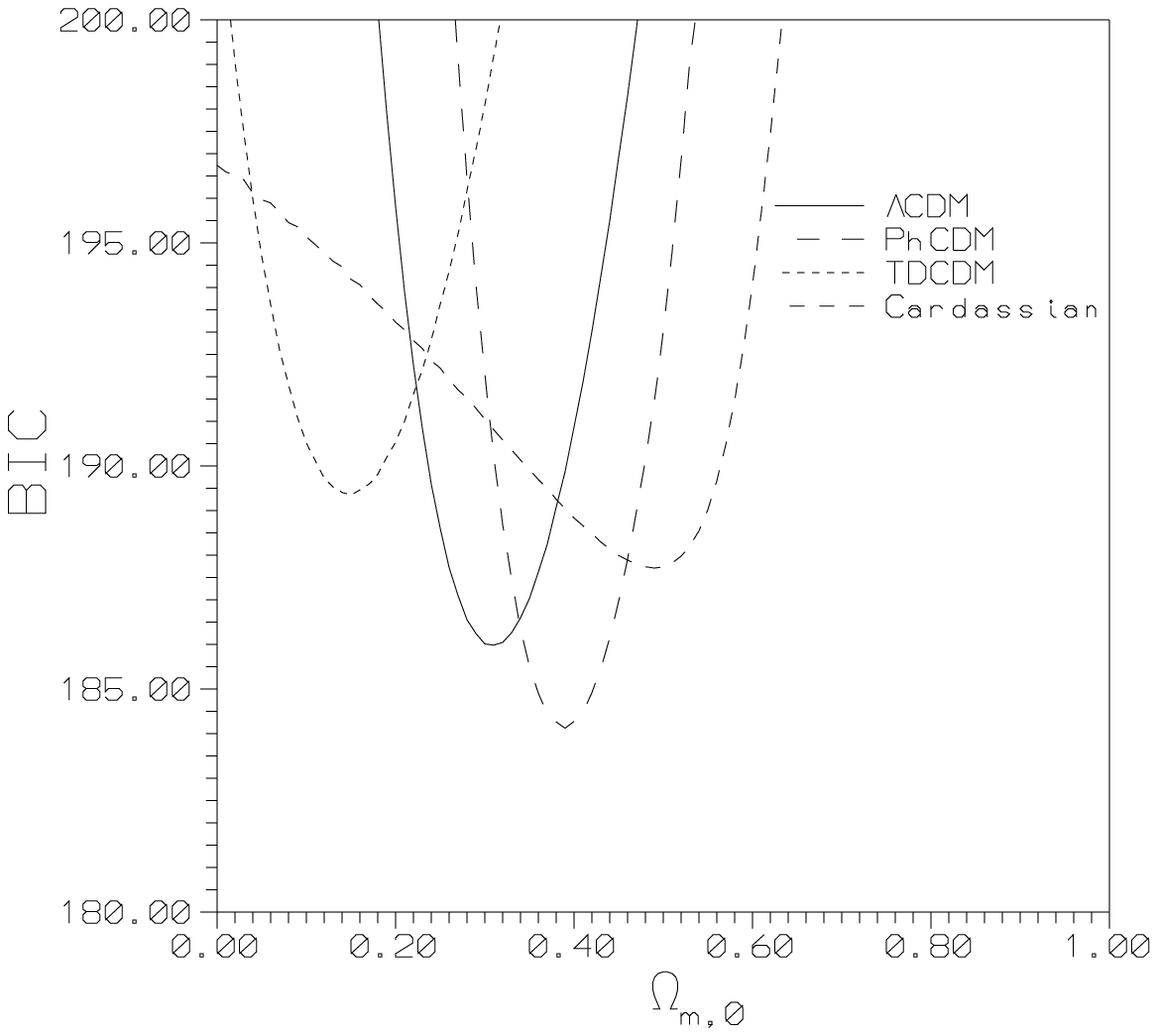}
\caption{The values of the BIC in respect to fixed value
$\Omega_{m,0}$ for four flat models (with topological defect, with
the cosmological constant, with phantom and for the Cardassian model).}
\label{fig:2}
\end{figure}

At first we analyze three flat models with two free parameters, i.e., the
flat $\Lambda$CDM, TDCDM and PhCDM models. There is a significant difference
between predictions of these models. The $\Lambda$CDM model prefers a universe
with $\Omega_{m,0}$ close to $0.3$, the PhCDM model favors a high
density universe while the TDCDM model favors a low density universe.
In Fig.~\ref{fig:1} and Fig.~\ref{fig:2} we present values of the AIC and BIC
for these and Cardassian models. If $\Omega_{m,0}<0.22$ then the
information criteria favor the TDCDM model. For $\Omega_{m,0} \in
(0.22, 0.34)$, the $\Lambda$CDM is favored while for $\Omega_{m,0}>0.34$
the PhCDM model is preferred. The AIC allows also the Chaplygin gas model.
However the BIC again prefers the $\Lambda$CDM model against the Chaplygin
gas model.
 
The similar analysis with the use of the information criteria is done in the
case of the assumed prior $\Omega_{m,0}=0.3$ \cite{Peebles:2002gy}
(Table~\ref{tab:4}). In this case the model with $\Lambda$ is preferred over
the models with phantoms, that is in contrary to the results obtained in the
previous case of no priors for $\Omega_{m,0}$. It clearly shows that
more precise measurements of $\Omega_{m,0}$ will give us the
possibility to discriminate between the $\Lambda$CDM and PhCDM models.
 
For the model models with the modified FRW equation information criteria
prefferes only the Cardassian model. However this model is preferred
in the case of a high density universe, with $\Omega_{m,0}>0.44$ which
seems to be too high with comparison with present extragalactic data
\cite{Peebles:2002gy}. 

\begin{figure}
\includegraphics[width=0.9\textwidth]{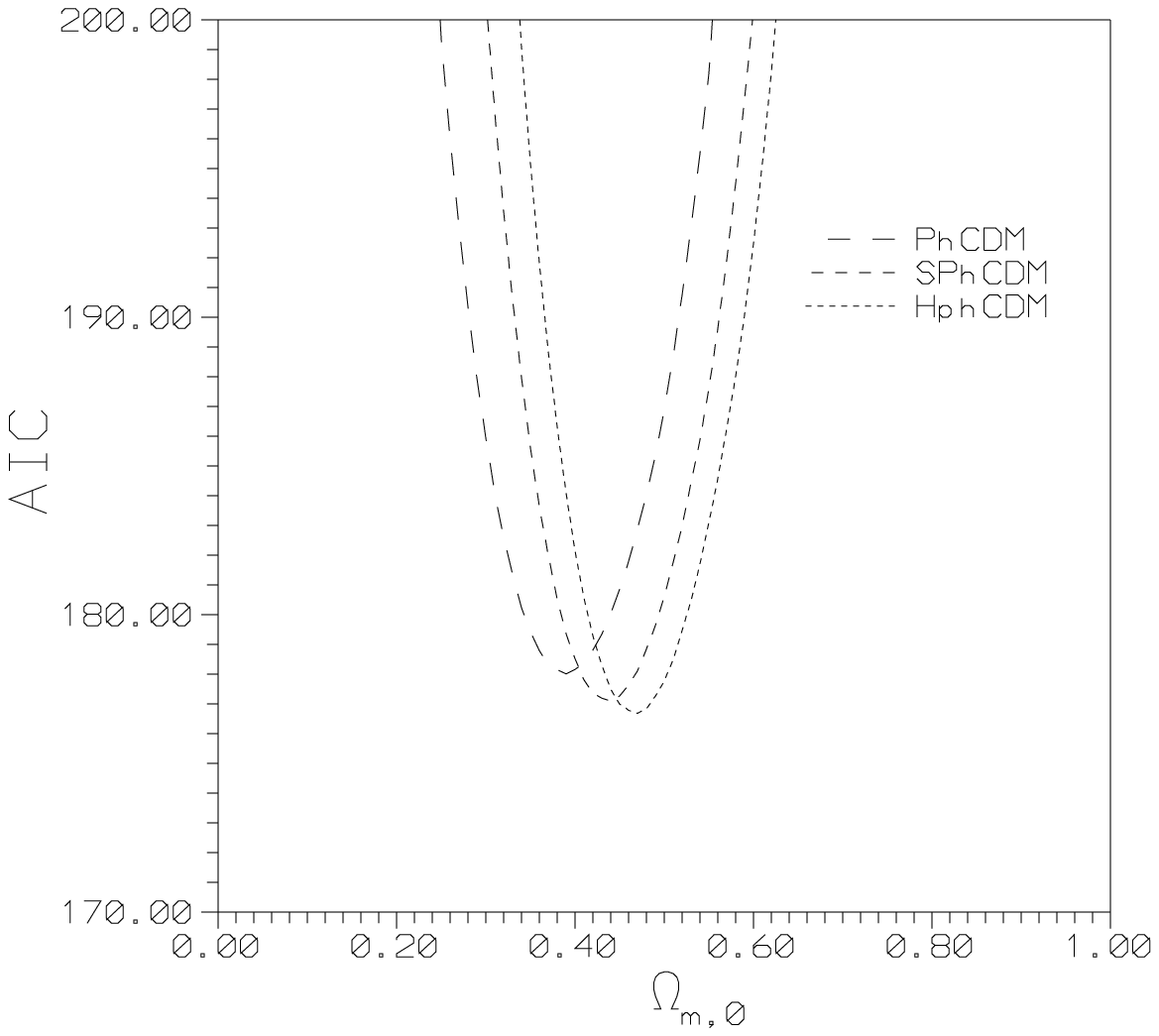}
\caption{The values of the AIC in respect to fixed value
$\Omega_{m,0}$ for three flat phantom, super nad hiper phantoms models 
with different fixed value of $w$ - PhCDM ($w=-4/3$), SPhCDM ($w=-5/3$),
HPhCDM ($w=-2$).}
\label{fig:3}
\end{figure}

\begin{figure}
\includegraphics[width=0.9\textwidth]{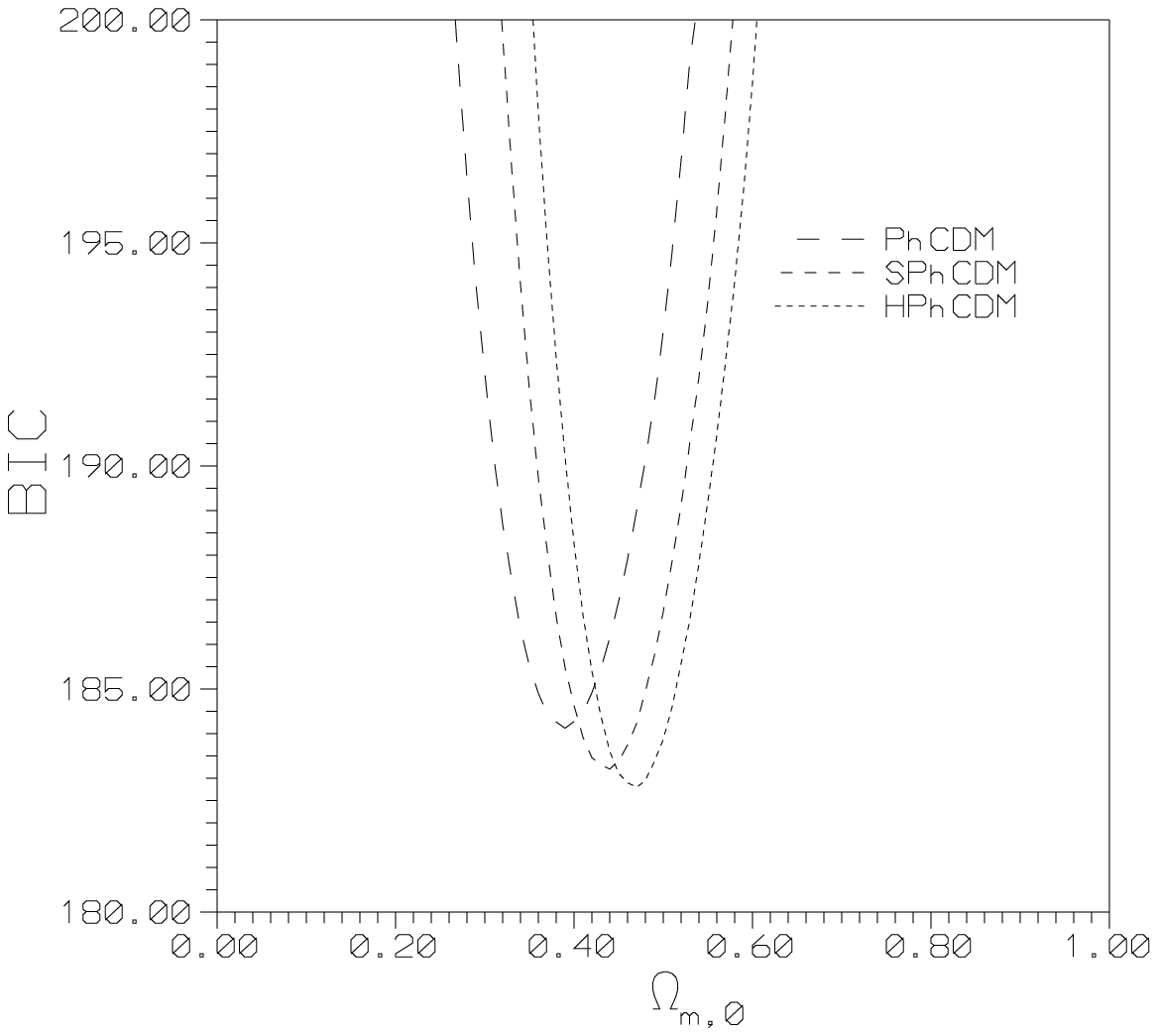}
\caption{The values of the BIC in respect to fixed value
$\Omega_{m,0}$ for three flat phantom, super nad hiper phantoms models 
with different fixed value of $w$ - PhCDM ($w=-4/3$), SPhCDM ($w=-5/3$), 
HPhCDM ($w=-2$).}
\label{fig:4}
\end{figure}

The comparison of PhCDM models with different fixed value of $w_X$ shows 
Fig.~\ref{fig:3} and Fig.~\ref{fig:4}. Note that it is required to have 
higher matter content to equilize the greater negative effect of lower 
(more negative) value of the factor $w$. The model which minimize the values 
of both AIC and BIC is the model with $w=-2.37$.

\begin{figure}
\includegraphics[width=0.9\textwidth]{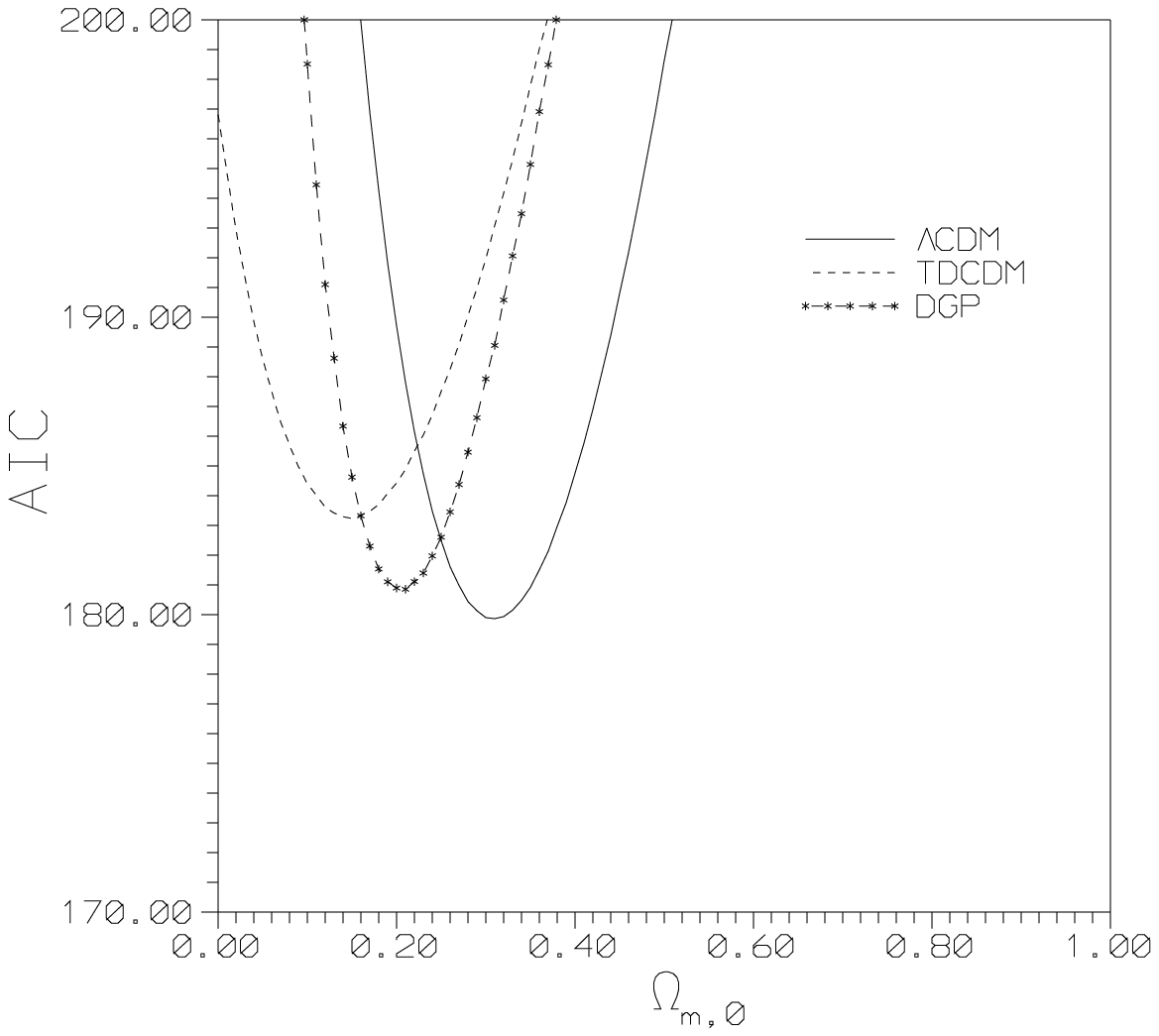}
\caption{The values of the AIC in respect to fixed value
$\Omega_{m,0}$ for low density (TDCDM and DGP) models and $\Lambda$CDM 
model.} 
\label{fig:5}
\end{figure}

\begin{figure}
\includegraphics[width=0.9\textwidth]{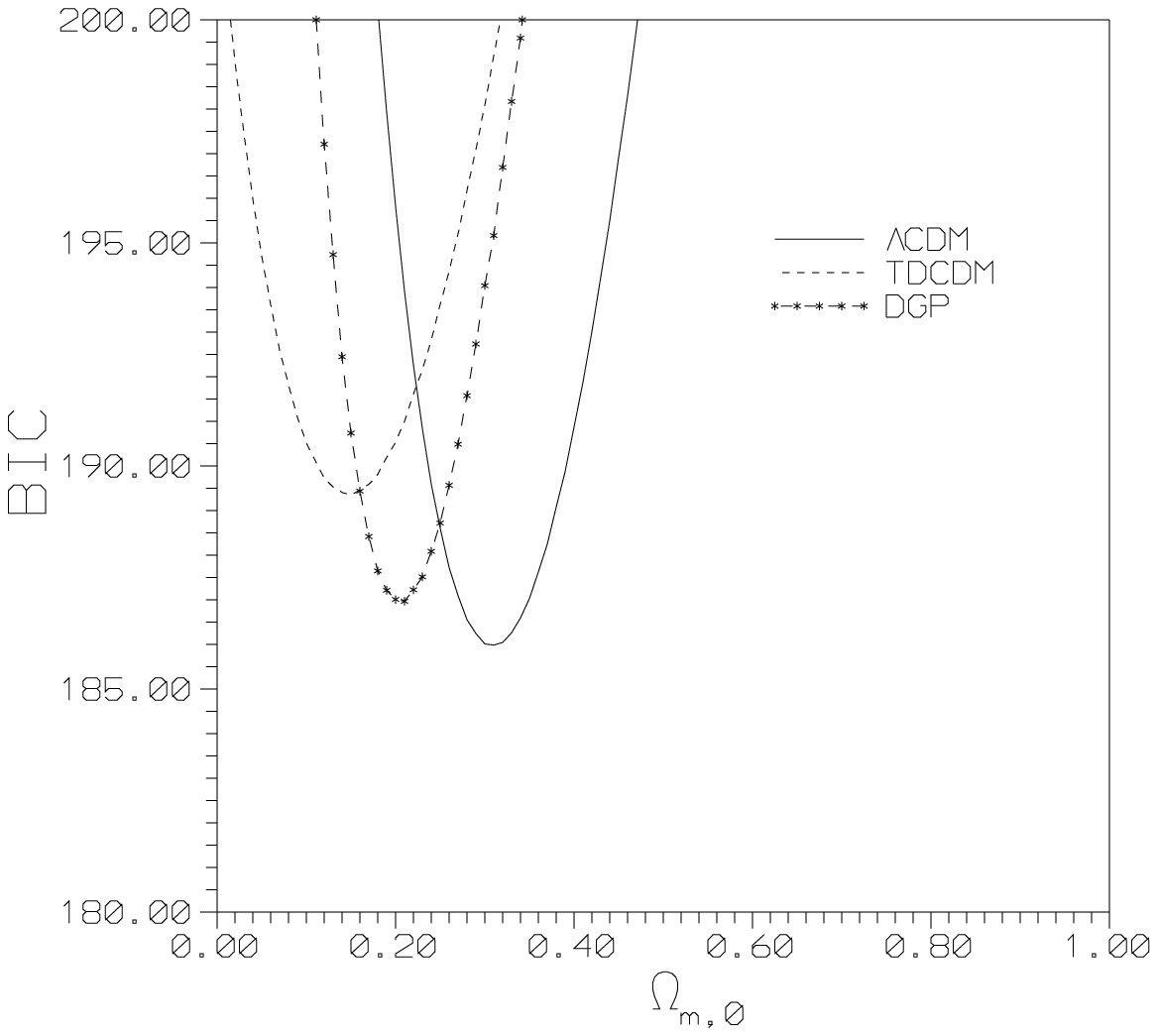}
\caption{The values of the BIC in respect to fixed value
$\Omega_{m,0}$ for low density (TDCDM and DGP) models and $\Lambda$CDM 
model.} 
\label{fig:6}
\end{figure}

Fig.~\ref{fig:5} and Fig.~\ref{fig:6} shows the AIC and BIC in respect to
$\Omega_{m,0}$, for low density universes (the TDCDM and DGP models)
and the $\Lambda$CDM model. Dependent on value of $\Omega_{m,0}$
the different cosmological models are selected. We find that TDCDM model
is distinguished for $\Omega_{m,0}<0.16$, DGP model for
$\Omega_{m,0} \in (0.16, 0.24)$ and $\Lambda$CDM for 
$\Omega_{m,0}>0.24$.
 
For deper analysis of statistical results it would be useful to consider
the information entropy of the distribution  which is defined as
\begin{equation}
\label{eq:13}
Entropy = - \Sigma f_i \log_a (f_i)
\end{equation}
where $a$ is a number of independent states of the system.
 
In Table~\ref{tab:7} the value of entropy for four flat models (with
topological defect, with the cosmological constant, with phantom
and for the Cardassian model). The value of entropy for one dimensional PDF
($\Omega_{m,0}$) is also presented. We can see that in both cases
we obtain minimal value of entropy for the PhCDM model.
 
Basing on these simple and objective information criteria and the minimum
entropy principle we obtain that SNIa data favor the models with dark energy
rather than based on modified FRW equations. Among the dark energy models the
best candidates are the PhCDM and $\Lambda$CDM models. However, to make the
final decision which model describes our Universe it is necessary to obtain
the precise value of $\Omega_{m,0}$ from independent observations.

\section{Conclusion}
 
The main goal of this paper is to decide which class of models: dark energy
or models based on the modified FRW equation are distinguished by statistical
analysis of SNIa data. For this aims we use the Akaike and Bayesian
information criteria.
 
It was considered two groups of five different models containing dust
and dark energy (the first group) and dust matter and an additional term which
modifies FRW dynamics (the second group). Our main conclusion is that both 
criteria weigh in favor of the flat dark energy models. We argue that there is
no strong reason to favor models with modified FRW dynamics over FRW dark energy 
models.
 
We are also able to decide which model with dark energy is distinguished by
the statistical analysis of SNIa data. To do this we use the Akaike and
Bayesian information criteria. The former criterion weighs in favor of the
flat phantom model while the latter distinguishes the flat phantom and
$\Lambda$CDM models. Assuming the prior $\Omega_{m,0}=0.3$ both the
AIC and BIC criteria weighs in favor of the model with dark energy, namely
the flat $\Lambda$CDM model.
 
The further conclusions are the following.
\begin{itemize}
\item The minimal number of essential parameters in the cosmological models
with dark energy is in principal two, i.e., ($H_0,\Omega_{m,0}$).
The list of essential parameters may be longer, because some of them are not
convincingly measured with present data, like the parameter $w_1$.
\item The curvature density parameter does not belong to the class of
essential parameters when all the rest parameters are without any priors
(with no fixed $\Omega_{m,0}$). At this point our result coincides
with analogous result obtained by Liddle who found it basing on other
observations.
\item If we consider models in which all model parameters are fitted then the
PhCDM model with double initial and final singularities is distinguished.
\item When we consider the prior on $\Omega_{m,0}$ then while
for $\Omega_{m,0} < 0.16$ the model with two-dimensional topological
defect is favored. The value  $\Omega_{m,0} \in (0.16;0.24)$ favor
DGP model while the value  $\Omega_{m,0} \in (0.22;0.34)$ favor
the $\Lambda$CDM model. For $\Omega_{m,0} \in (0.22;0.34)$ the phantom
model is preferred. With high density Universe ($\Omega_{m,0}>0.44)$ Cardassian
model (with modified FRW equation of state) is preferred.
\end{itemize}
 
During our analisis we have use recently avaliable SNIa data and of course 
future SNAP data will provide mor sophisticated information to distinguish 
between FRW dark energy models and the models with modified FRW dynamics.

To make the ultimate decision which model describes our Universe it is
necessary to obtain the precise value of $\Omega_{m,0}$ from
independent observations.

\ack{Authors thanks dr A.Krawiec for helpful discussion.
M. Szydlowski acknowledges the support by KBN grant no. 1 P03D 003 26.}


\begin{table}
\noindent
\caption{The values of the AIC and BIC for the models from
Table \ref{tab:1} both for flat and non-flat cases.}
\label{tab:3}
\begin{tabular}{c|cccc}
\hline \hline
case & AIC ($\Omega_{k,0}=0$) & AIC ($\Omega_{k,0} \ne 0$)
& BIC ($\Omega_{k,0}=0$) & BIC ($\Omega_{k,0} \ne 0$) \\
\hline
0   & 325.5  & 194.4 & 328.6  & 200.5  \\
1   & 179.9  & 179.9 & 186.0  & 189.0  \\
2   & 183.2  & 180.1 & 189.4  & 194.4  \\
3a  & 178.0  & 179.3 & 184.1  & 188.5  \\
3b  & 178.5  & 179.7 & 187.7  & 191.9  \\
4a  & 179.7  & 181.4 & 188.9  & 193.6  \\
4b  & 181.7  & 183.4 & 193.9  & 198.7  \\
5a  & 180.5  & 182.0 & 192.7  & 197.3  \\
5b  & 180.4  & 181.9 & 192.6  & 197.2  \\
\hline
\end{tabular}
\end{table}

\begin{table}
\noindent
\caption{The values of the AIC and BIC for the models from
Table \ref{tab:2} both for flat and non-flat cases.}
\label{tab:4}
\begin{tabular}{c|cccc}
\hline \hline
case & AIC ($\Omega_{k,0}=0$) & AIC ($\Omega_{k,0} \ne 0$)
& BIC ($\Omega_{k,0}=0$) & BIC ($\Omega_{k,0} \ne 0$) \\
\hline
1   & 180.9  & 180.0 & 187.0  & 189.1  \\
2a  & 327.0  & 184.6 & 336.2  & 196.8  \\
2b  & 182.1  & 183.8 & 194.3  & 199.1  \\
3   & 178.5  & 179.7 & 187.7  & 191.9  \\
4   & 183.9  & 183.6 & 196.2  & 198.8  \\
5a  & 315.1  & 195.8 & 321.2  & 205.0  \\
5b  & 180.3  & 181.6 & 189.4  & 193.8  \\
\hline
\end{tabular}
\end{table}
 
\begin{table}
\noindent
\caption{The values of the AIC and BIC for the models from
Table \ref{tab:1} with the prior $\Omega_{m,0}=0.3$ both for flat
and non-flat cases.}
\label{tab:5}
\begin{tabular}{c|cccc}
\hline \hline
case & AIC ($\Omega_{k,0}=0$) & AIC ($\Omega_{k,0} \ne 0$)
& BIC ($\Omega_{k,0}=0$) & BIC ($\Omega_{k,0} \ne 0$) \\
\hline
0   &  ---   & 216.9 &  ---   & 220.0  \\
1   & 177.9  & 179.9 & 181.0  & 186.0  \\
2   & 190.0  & 178.8 & 193.0  & 184.9  \\
3a  & 183.9  & 179.6 & 187.0  & 186.7  \\
3b  & 179.9  & 178.2 & 186.0  & 187.4  \\
4a  & 179.6  & 179.8 & 185.7  & 188.9  \\
4b  & 181.6  & 181.8 & 190.8  & 194.0  \\
5a  & 179.2  & 180.2 & 188.4  & 192.4  \\
5b  & 178.8  & 180.3 & 187.9  & 192.5  \\
\hline
\end{tabular}
\end{table}
 
\begin{table}
\noindent
\caption{The values of the AIC and BIC for models from
Table~\ref{tab:2} with the prior $\Omega_{m,0}=0.3$ both for flat and
non-flat cases.}
\label{tab:6}
\begin{tabular}{c|cccc}
\hline \hline
case & AIC ($\Omega_{k,0}=0$) & AIC ($\Omega_{k,0} \ne 0$)
& BIC ($\Omega_{k,0}=0$) & BIC ($\Omega_{k,0} \ne 0$) \\
\hline
1   & 185.9  & 178.2 & 189.0  & 184.3  \\
2a  & 474.7  & 183.0 & 480.9  & 192.2  \\
2b  & 180.5  & 181.9 & 189.7  & 194.1  \\
3   & 179.9  & 178.2 & 186.0  & 187.4  \\
4   & 181.9  & 183.9 & 191.1  & 196.1  \\
5a  &1296.3  & 204.4 &1299.3  & 210.5  \\
5b  & 179.5  & 180.1 & 185.6  & 189.3  \\
\hline
\end{tabular}
\end{table}

\begin{table}
\noindent
\caption{The value of information entropy for four flat models (with
topological defect, the cosmological constant, phantom and for the Cardassian
model). The value of entropy for the one dimensional PDF of
$\Omega_{m,0}$ is also presented.}
\label{tab:7}
\begin{tabular}{c|cc}
\hline \hline
model & entropy & entropy ($\Omega_{m,0}$) \\
\hline
$\Lambda$CDM   & 0.604  & 0.601   \\
TdCDM         & 0.627  & 0.641   \\
PhCDM         & 0.591  & 0.577   \\
Cardassian    & 0.718  & 0.682   \\
\hline
\end{tabular}
\end{table}


\end{document}